\documentclass[epj,nopacs]{svjour}
\usepackage{amsmath}
\usepackage{amssymb}
\usepackage{latexsym}
\usepackage[dvips]{graphicx}
\usepackage[english]{babel}

\def\bge{\begin{equation}}
\def\ene{\end{equation}}
\def\bgea{\begin{eqnarray}}
\def\enea{\end{eqnarray}}

\def\bge{\begin{equation}}
\def\ene{\end{equation}}
\def\bgea{\begin{eqnarray}}
\def\enea{\end{eqnarray}}

\def\ls{\raise 1.5pt\hbox{$\,<\;$}\kern -10.5pt\lower3.5pt
          \hbox{$\sim$}\kern 1.5pt} %%% less or similar
\def\gs{\raise 1.5pt\hbox{$\,>\,$}\kern -9.5pt\lower3.5pt
          \hbox{$\sim$}\kern 1.5pt} %%% greater or similar

\usepackage{color}

\usepackage{ulem} % use: \sout{Striked out text}

\begin{document}
\sloppy
\title{A Solution to the electroweak horizon problem \\
in the $R_{\rm h}=ct$ universe}
\author{Fulvio Melia\thanks{John Woodruff Simpson Fellow.}}
\institute{Department of Physics, the Applied Math Program, and Department of Astronomy, \\
              The University of Arizona, Tucson, AZ 85721,
              \email{fmelia@email.arizona.edu}}

\authorrunning{Melia}
\titlerunning{Electroweak horizon problem}

%\date{\today}
%\date{March 10, 2010}
%\date{May 07, 2010}
\date{August 8, 2018}
%\date{}

\abstract{Particle physics suggests that the Universe may have undergone several
phase transitions, including the well-known inflationary event associated
with the separation of the strong and electroweak forces in grand unified
theories. The accelerated cosmic expansion during this transition, at 
cosmic time $t\sim 10^{-36}-10^{-33}$ seconds, is often viewed as 
an explanation for the uniformity of the CMB temperature, $T$, which 
would otherwise have required inexplicable initial conditions. With the 
discovery of the Higgs particle, it is now quite likely that the Universe
underwent another (electroweak) phase transition, at $T=159.5\pm1.5$ 
GeV---roughly $\sim 10^{-11}$ seconds after the big bang. During this
event, the fermions gained mass and the electric force separated from the 
weak force. There is currently no established explanation, however, for the 
apparent uniformity of the vacuum expectation value of the Higgs field which, 
like the uniformity in $T$, gives rise to its own horizon problem in standard 
$\Lambda$CDM cosmology. We show in this paper that a solution to the 
electroweak horizon problem may be found in the choice of cosmological model, 
and demonstrate that this issue does not exist in the alternative 
Friedmann-Robertson-Walker cosmology known as the $R_{\rm h}=ct$ universe.}
%    \PACS{{04.20.Ex},\  {95.36.+x},\  {98.80.-k},\  {98.80.Jk}}
\maketitle

% ---------------------------------------------------------------------------------
\section{Introduction}
Several phase transitions in particle physics have potentially
deep implications for cosmology. A well known example is the 
phase transition associated with grand unified theories (GUT), during
which the strong and electroweak (EW) forces are believed to have
separated. This well-studied case was originally motivated by missing 
magnetic monopoles, but was quickly identified as an inflationary event 
\cite{Guth:1981} that could solve the horizon problem in standard 
$\Lambda$CDM cosmology. 

Today, with the discovery of the Higgs particle \cite{Aad:2012}, the 
consequences of a second well-motivated transition---the electroweak 
phase transition (EWPT)---occurring at a critical temperature of
$159.5\pm 1.5$ GeV, are being studied with increasing interest.
In $\Lambda$CDM, this temperature would have been reached at cosmic
time $t\sim 10^{-11}$ seconds, well past the first (inflationary)
transition at $t\sim 10^{-36}-10^{-33}$ seconds. The standard model 
EWPT is now known to be a `crossover,' i.e., one that does not depart 
far from equilibrium, rather than first order (with a discontinuity), 
that would have provided a ready explanation for the origin of baryon 
asymmetry, i.e., the matter left over after the annihilations between
matter and anti-matter ended at very early times. But many extensions 
to the standard model of particle physics allow additional Higgs 
fields that reopen the possibility of a first-order phase transition at 
the EW scale \cite{Fileviez:2009} which would, in addition, generate 
gravitational waves (e.g., \cite{Weir:2018}) measurable with LISA and other 
next-generation detectors \cite{Caprini:2016,Audley:2017}. 
Learning about a possible first-order EWPT by measuring the Higgs 
self-interaction will also be a goal of the High-Luminosity Large Hadron 
Collider (LHC) and other future colliders (see, e.g., 
\cite{Noble:2008,Dolan:2012,Barr:2015}). Of course, first and foremost, 
the crucial function of the EWPT is the generation of fermionic mass and 
the consequent separation of the electric and weak forces, both crucial
events in the history of the Universe.

As the Universe cooled down further following the EWPT, a third phase 
transition is believed to have occurred at roughly $100$ MeV, corresponding
to a time $t\sim 10^{-6}$ seconds in $\Lambda$CDM. This would have arisen out
of quantum chromodynamics, associated with the transformation of quarks 
behaving like free particles (in a quark-gluon plasma at asymptotically
high temperatures) into the `confined states' of baryons and mesons in 
the hadronic phase as the Universe continued to expand. 

Our focus in this paper is the EWPT because, as we shall see,
the nature of the Higgs field appears to lead inevitably to yet another 
horizon problem, not unlike what happened with the CMB temperature, 
though this time having to do with the vacuum expectation value (vev) of the 
Higgs field, which appears to be universal---even on scales exceeding 
causally-connected regions. Inflation was invoked to account for the 
uniformity of the CMB on large scales, but the accelerated expansion it 
spawned would have occurred well {\it before} the EWPT, and would therefore 
have been largely irrelevant to the Higgs vev. 

There is no well-established solution yet to this so-called electroweak horizon 
problem (EHP), which has been recognized in various forms over the past half 
century. At first, the inclination was to search for sub-horizon features
produced in the EWPT. For example, Zel’dovic, Kobzarev \& Okun 
\cite{Zeldovic:1975} and Kibble \cite{Kibble:1976} 
offered an early assessment of the possibility that domain 
walls might have been created in the cosmos as a result of such scale 
transitions in the early Universe. These topological defects would have 
significant observational consequences, e.g., producing measurable anisotropies 
in the CMB temperature \cite{Vilenkin:1994,Lazanu:2015,Sousa:2015}. 
As cosmological observations have improved, 
however, it has become increasingly clear that the most likely resolution 
of the EHP is to avoid it in the first place. More recent attempted solutions 
have therefore included a late-time weak-scale inflation 
\cite{Randall:1995,Lyth:1996,Randall:1996,German:2001,Boeckel:2010,Davoudiasl:2016}, 
though no particular proposal has had any impact with our 
interpretation of the observations thus far.

In this paper, we suggest that the EHP may be due to an incorrect
choice of the cosmology, and propose that the solution may be found---not in 
a tweaked $\Lambda$CDM but, rather---in the alternative Friedmann-Robertson-Walker 
(FRW) cosmology known as the $R_{\rm h}=ct$ Universe 
\cite{Melia:2007,MeliaShevchuk:2012,Melia:2016a,Melia:2017a}. 
We shall demonstrate that, just as $R_{\rm h}=ct$
avoids the horizon problem with the CMB temperature \cite{Melia:2013a}, it is equally 
free of any subsequent horizon problem with the EWPT. The critical difference 
between $\Lambda$CDM and $R_{\rm h}=ct$ that allows this to happen is that, 
while the former has an early decelerated expansion, the latter does not.

\section{Background}
The Higgs mechanism for generating fermionic mass is now widely accepted 
\cite{Englert:1964,Higgs:1964}. There are actually two parts to this 
story: the first has to do with when (and if) a non-zero vev 
is acquired (which particle physicists commonly refer to as `turning on the 
Higgs field'); the second has to do with the size of the coupling constants 
with which the various elementary particles sense the Higgs field. At 
asymptotically high temperatures, the EW symmetry is unbroken. In simple 
terms, this means that all the `messenger' particles carrying the 
electroweak force transfer the same amount of momentum per unit energy 
from one fermion to the next. In this regime, the relativistic expression 
for energy, $E^2=m^2c^4+p^2c^2$, does not differentiate among them based 
on the value of $p/E$ because they all have $m=0$, regardless of whether 
the particle is a photon, a $W^\pm$ or a $Z$. 

The electric and weak forces separate, however, when $p/E$ changes due to 
the emergence of a non-zero mass. This `spontaneous symmetry breaking'
happens only when the Higgs acquires a non-zero vev and the particles
have unequal coupling constants, so that their masses are different. But 
note that even if the Higgs mechanism did not exist, symmetry would be 
attained at asymptotically high temperatures anyway, because in that
limit $E/mc^2 \gg 1$, which would make the ratio $p/E$ virtually identical 
for all the bosons. The symmetry would still have been broken as the 
temperature dropped, as long as their inertial masses were different. 
The viability of the Higgs mechanism just makes the spontaneous symmetry 
breaking cleaner and more precisely localized in temperature---and 
therefore redshift, or cosmic time. As noted in the introduction, we 
now know that the EWPT must have occurred at the critical temperature
$T=159.5\pm1.5$ GeV, when $t\sim 10^{-11}$ seconds in $\Lambda$CDM.

But what sets the Higgs vev? As of today, there is no known 
theoretical constraint on this important property of the Higgs field.
This critical temperature could have been something else. One may
reasonably argue, however, that whatever conditions establish the vev,
it is manifested uniformly throughout a causally-connected 
region of spacetime. There is no reason, though, why the same vev 
should emerge everywhere, even at distances exceeding an observer's
causal horizon. 

Donoghue et al. \cite{Donoghue:2010} took an interesting approach to this question
by estimating the likelihood function for the Higgs vev based on
anthropic constraints on the existence of atoms. It is known that
nuclei and atoms would not exist if the masses of light quarks and
the electron were modestly different from their measured values
\cite{Agrawal:1998,Hogan:2000,Damour:2008}. And
since the fermionic masses are proportional to the Higgs vev,
these anthropically permitted bounds may be interpreted as constraints
on the Higgs vev distribution accessible to us, given the other
parameters in the standard cosmological model. 

They explored how the Higgs vev distribution function is shaped by
possible variations in the cosmology, always with the constraint that
nuclei and atoms should appear. This assumes, of course, that there
exists an a priori range of vev's, based on an unknown property
of the fundamental theory. Particle physicists expect that the
Higgs vev can take on any value throughout a very large domain,
extending at least up to the GUT scale, many orders
of magnitude above the EW scale (a disparity in energy known as 
the `hierarchy problem').

They found that even within an anthropic framework, there is no
reason to expect a single, observed uniform Higgs 
vev, commonly referred to as $v_0$. The distribution estimated in this 
fashion peaks near $v_0$, though it extends over several orders of 
magnitude. Its median value is actually $2.25v_0$, and its $2\sigma$ 
range extends from $0.10v_0$ to $11.7v_0$. Recalling that fermionic
masses are proportional to the vev, we therefore see that nuclear
and atomic properties could in principle have varied by at least
one to two orders of magnitude across the Universe, from one
causally-connected region to another. Yet no such variation has
ever been confirmed.

\section{The Electroweak Horizon Problem}
To understand why this is a problem with standard cosmology,
let us examine the size of a causally-connected region
within which one may expect to find a uniform Higgs vev
(see fig.~1). In ref.~\cite{Melia:2013a}, we showed that the null 
geodesic equation for a flat Universe may be written
\begin{equation}
\dot{R}_\gamma=c\left({R_\gamma\over R_{\rm h}}-1\right)\;,
\end{equation}
where $R_{\rm h}\equiv c/H(t)$ is the Hubble (or gravitational)
radius in terms of the Hubble parameter $H(t)$ \cite{Melia:2007,MeliaShevchuk:2012}, 
and $R_\gamma$ is the proper radius 
of a photon propagating along the null geodesic reaching the 
observer at $R_\gamma=0$ (point B in this figure). In the FRW 
framework, a proper radius may also be written as $R\equiv a(t)r$, 
in terms of the comoving radius $r$ and the universal expansion 
factor $a(t)$. When the cosmic equation of state is written in 
the form $p=w\rho$, for the total pressure $p$ and energy density 
$\rho$, the gravitational radius satisfies the dynamical equation 
\cite{MeliaAbdelqader:2009}
\begin{equation}
\dot{R}_{\rm h}={3\over 2}(1+w)c\;.
\end{equation}
For example, in a radiation dominated universe, with $w=1/3$,
the Hubble (or gravitational) radius expands at twice the speed
of light.

Solving Equations~(1) and (2) simultaneously yields the null
geodesic $R_\gamma(t)$ linking a source emitting photons
at $R_{\rm src}(t_e)=R_\gamma(t_e)$ at time $t_e$, with the
observer who receives them at $R_\gamma(t_o)=0$ at time $t_o$. 
Note that for a given observer at time $t_o$, there is a unique
(radial) null geodesic arriving at his location (see fig.~2). 
This function $R_\gamma(t)$ begins at $R_\gamma(0)=0$ at time 
$t=0$ (i.e., the big bang), increases while $R_\gamma>R_{\rm h}$ 
and reaches a maximum at $t_{\rm max}$ defined by the condition 
$R_\gamma(t_{\rm max})=R_{\rm h}(t_{\rm max})$, and then 
decreases again towards $R_\gamma(t_o)=0$ once $R_{\rm h}$ 
overtakes $R_\gamma$ so that $\dot{R}_\gamma<0$ in Equation~(1) 
(see ref.~\cite{Melia:2013a}). 

For a broad range of conditions, previous studies have shown 
that an observer receiving light at time $t$ ($>t_{\rm max}$)
sees a maximum photon excursion $R_{\gamma{\rm o}}(t_{\rm max})\lesssim 
R_{\rm h}(t)/2$ away from his position \cite{Bikwa:2012,Melia:2012,Melia:2013b,Melia:2018}. 
As explained more extensively in ref.~\cite{Melia:2013b}, 
this behaviour of null geodesics in FRW is not
difficult to understand. All models other than de Sitter had
no pre-existing detectable sources away from the observer's
location before the big bang. The photons we detect at time 
$t$ from the most remote distances were emitted only {\it after} 
their sources had sufficient time to reach these extreme 
locations, which lie at roughly half of $R_{\rm h}(t)$. 
Therefore the proper size of our {\it visible} Universe at 
any given time $t$ is only about half of the Hubble (or 
gravitational) radius $R_{\rm h}(t)$ (see also ref.~\cite{Melia:2018}).  
Claims that we see sources today (at time $t_0$) beyond 
$R_{\rm h}(t_0)$ (see, e.g., ref.~\cite{Davis:2001}) are 
simply confusing where the sources are today with where they 
were when they emitted the light we are receiving now. Our 
causally-connected region is based solely on the proper size 
(i.e., $R_\gamma[t_{\rm max}]$) of the volume within which 
light signals have been exchanged. Photons radiated by sources 
beyond $R_{\rm h}$ may be detectable in our future, but 
they have no relevance to the causally connected spacetime 
points today. 

We shall discuss the implications of the null geodesics shown
in fig.~2 for $R_{\rm h}=ct$ shortly, but first we examine
why an EWPT horizon problem emerges in $\Lambda$CDM. We shall write 
the Hubble parameter for flat $\Lambda$CDM cosmology in the form
\begin{equation}
H(a)=H_0\sqrt{\Omega_{\rm m}\,a^{-3}+\Omega_{\rm r}\,a^{-4}+\Omega_\Lambda}\;,
\end{equation}
where today's Hubble constant ($H_0=67.8$ km s$^{-1}$ Mpc$^{-1}$) and 
the scaled densities for matter ($\Omega_{\rm m}=0.308$), radiation 
($\Omega_{\rm r}=5.37\times 10^{-5}$) and dark energy 
($\Omega_\Lambda=1-\Omega_{\rm m}-\Omega_{\rm r}$), are assumed to
have their {\it Planck} values \cite{Planck:2015}. The redshift and age at
decoupling are, respectively, $z_{\rm cmb}=1089.9$ and $t_{\rm cmb}=377,700$
years. Therefore, the expansion factor at this epoch was
$a(t_{\rm cmb})=(1+z_{\rm cmb})^{-1}\approx 9.17\times 10^{-4}$,
implying a Hubble constant $H(t_{\rm cmb})\approx 4.78\times 10^{-14}$ s$^{-1}$.
The Hubble (gravitational) radius at decoupling was therefore
$R_{\rm h}(t_{\rm cmb})=c/H(t_{\rm cmb})\approx 0.20$ Mpc.

Integrating the equation
\begin{equation}
t_{\rm cmb}-t=\int_a^{a_{\rm cmb}}{da\over a\,H(a)}\;,
\end{equation}
it is straightforward to find $a(t)$ and $H(t)$ with the
use of Equation~(3) at any time $t<t_{\rm cmb}$, since we
terminate the calculation at $t_{\rm ew}=10^{-11}$ seconds,
well after the inflationary phase ended at $t_f\sim 10^{-33}$
seconds. At the EWPT, one finds an expansion factor
$a(t_{\rm ew})\approx 1.93\times 10^{-4}$ and a corresponding
Hubble (gravitational) radius $R_{\rm h}(t_{\rm ew})\approx
0.016$ Mpc.

\begin{figure}
\begin{center}
\includegraphics[width=\linewidth]{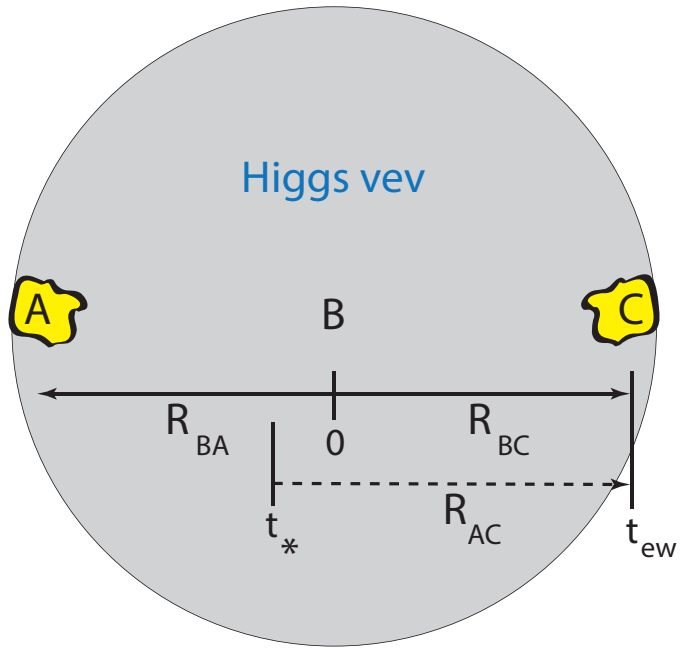}
\end{center}
\caption{Schematic diagram showing observer B connecting
causally with two opposite patches (A and C) in the Higgs
vev, a proper distance $R_{\rm BA}(t)=R_{\rm BC}(t)$ away.
Patch A emitted a light signal at time $t_*$ that reached C,
a proper distance $R_{\rm AC}(t_{\rm ew})$ away, at the
electroweak phase transition time $t_{\rm ew}$.}
\end{figure}

Solutions to the geodesic Equation~(1) therefore suggest that 
the size of a causally-connected region at $t_{\rm ew}$ would have
been $R_{\rm ew}(t_{\rm ew})\lesssim R_{\rm h}(t_{\rm ew})/2\approx 0.008$ Mpc,
and shifting forward to today, this scale expands to 
$R_{\rm ew}(t_0)\sim [a(t_0)/a(t_{\rm ew})]R_{\rm ew}(t_{\rm ew})\sim 41.5$ Mpc.
This proper radius would represent the size of the largest region
we should expect to see with a uniform Higgs vev today, yet we also
know from Equation~(1) that the proper size of our visible Universe 
right now is $\lesssim R_{\rm h}(t_0)/2\approx 2,212$ Mpc---more than
$50$ times bigger. If our understanding of the EWPT is correct, we
should therefore be seeing a variation of fermionic and atomic properties
across the Universe, which is absolutely not the case. This is the
electroweak horizon problem. And to amplify this point, recall that
one of the `hopes' for the EWPT is that it may explain the baryon
asymmetry. But this mechanism would not work because an $R_{\rm ew}(t_0)$
much smaller than $R_{\rm h}(t_0)$ would mean we should see pockets
of antimatter at proper distances exceeding the electroweak horizon,
and we simply have no evidence of such exotic domains.

\begin{figure}
\begin{center}
\includegraphics[width=\linewidth]{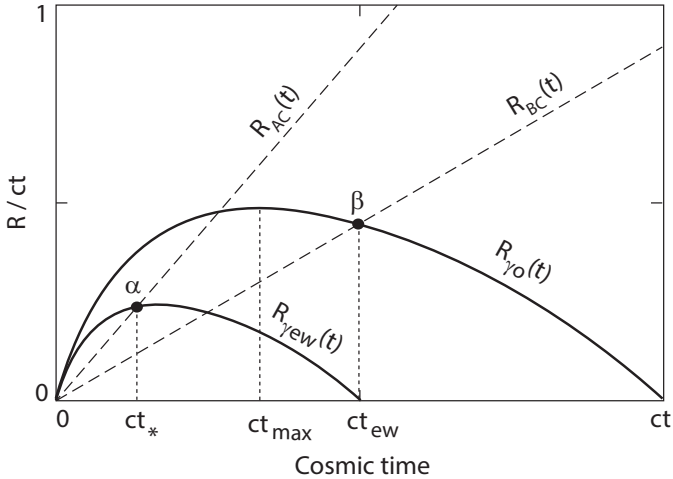}
\end{center}
\caption{Photon trajectories as seen by an observer in patch
C (see fig.~1) for the $R_{\rm h}=ct$ universe. The curve 
$R_{\gamma{\rm ew}}(t)$ represents the null geodesic arriving 
at C at the electroweak phase transition time $t_{\rm ew}$ 
from any source emitting light at proper distance 
$R_{\rm src}(t)=R_{\gamma{\rm ew}}(t)$ at time $t<t_{\rm ew}$. 
The other solid curve $R_{\gamma{\rm o}}(t)$ is the null 
geodesic for light arriving at C at time $t>t_{\rm ew}$.
The dashed lines $R_{\rm AC}(t)$ and $R_{\rm BC}(t)$ show the
proper distance from A to C and from B to C, respectively. 
The other labels and symbols are defined in the text.}
\end{figure}

As far as we can tell observationally, the Higgs vev is the same
throughout our visible Universe, even on opposite sides of us, with 
a separation more than 50 times larger than the electroweak horizon. 
As noted earlier, there is currently no established explanation for
how this could happen. This is the principal reason we are proposing
in this paper that the emergence of such a major hurdle with
the standard $\Lambda$CDM model is yet more evidence of a fundamental
problem with the cosmology, not the particle physics. In the next
section, we shall see that the alternative FRW cosmology known
as the $R_{\rm h}=ct$ universe has neither a CMB horizon problem,
nor a horizon problem with the EWPT. 

\section{A Solution of the EHP in $R_{\rm h}=ct$}
Our goal here is to understand how and why observer B sees a uniform
Higgs vev throughout his visible universe at time $t$, which necessitates
a causal connection between patches A and C by the time ($t_{\rm ew}$) 
the electroweak phase transition occurred. We shall find it easier to
describe the events from the perspective of an observer in C, however.
This won't affect how the proper distances are calculated because
$t$ is the same everywhere and, given the evident symmetry, C receives 
a signal from B at the same time that B receives a signal from C. 

In fig.~2 we show two relevant null geodesics and the worldlines
of patch A and observer B as seen by C (see fig.~1). The vertical axis 
gives the proper distance as a fraction of $ct$ for C receiving light 
signals at cosmic time $t_{\rm ew}$ and $t$. The label `$t_{\rm max}$' 
refers to the time at which the null geodesic $R_{\gamma 0}(t)$ attains
its maximum proper distance. A light signal is emitted by A at time
$t_*$ (labeled $\alpha$ in fig.~2), traveling towards C along the
trajectory $R_{\gamma{\rm ew}}(t)$. As long as these photons reach
C before a subsequent signal is emitted by B at $t_{\rm ew}$ (labeled
$\beta$), A and C will have been causally connected at the EWPT.

For a given cosmology, in this case $R_{\rm  h}=ct$, the null geodesic
$R_{\gamma 0}(t)$ is unique and, for a given time $t_{\rm ew}$, there
exists a single point $\beta$ satisfying the necessary conditions for
C to receive the signal from B at $t$. Turning this around, from B's 
perspective this happens identically from two opposite sides in the sky, 
and therefore $R_{\rm AC}(t_{\rm ew})=2R_{\rm BC}(t_{\rm ew})$. The
key question is thus whether there exists a time $t_*$ such that
A and C were causally connected at $t_{\rm ew}$, with a proper distance
$R_{\rm AC}(t_{\rm ew})$ that grew to fill the entire visible Universe
by time $t_0$ today.

In the $R_{\rm h}=ct$ universe, the expansion factor is $a(t)\propto t$
for all cosmic time. Therefore,
\begin{equation}
R_{\rm AC}(t_{\rm ew})=a(t_{\rm ew})\int_{t_*}^{t_{\rm ew}}c{dt^\prime\over
a(t^\prime)}=ct_{\rm ew}\,\ln(t_{\rm ew}/t_*)\;.
\end{equation}
Similarly,
\begin{equation}
R_{\rm BC}(t_{\rm ew})=a(t_{\rm ew})\int_{t_{\rm ew}}^{t_0}c{dt^\prime\over
a(t^\prime)}=ct_{\rm ew}\,\ln(t_0/t_{\rm ew})\;.
\end{equation}
It is trivial to verify that both of these proper distances satisfy the
null geodesic Equation~(1). The constraint $R_{\rm AC}(t_{\rm ew})=
2R_{\rm BC}(t_{\rm ew})$ therefore yields the condition
\begin{equation}
t_*=t_{\rm ew}\left({t_{\rm ew}\over t_0}\right)^2\;.
\end{equation}
Thus, no matter when the EWPT occurred relative to $t_0$, there is
always a time $t_*>0$---no matter how small---at which an exchange
of signals between patches A and C could have been initiated to 
ensure that they were causally connected by the time the Higgs 
vev was manifested. Put another way, regardless of how large $R_{\rm AC}$
is today ($t_0$), and regardless of when the EWPT took place ($t_{\rm ew}$), 
there always exists a physically meaningful value of $t_*$ that permitted 
A and C to be causally connected before the Higgs vev was imprinted 
on the cosmic structure observed by observer B at time $t_0$.

In this cosmology, the size of a causally-connected region at 
$t_{\rm ew}$ would have been $R_{\rm ew}(t_{\rm ew})\lesssim
ct_{\rm ew}/2\approx 0.3$ cm, much smaller than the corresponding
size in $\Lambda$CDM. But the critical difference between these
two models is that, whereas $\Lambda$CDM underwent significant
deceleration prior to $z\sim 0.7$, $R_{\rm h}=ct$ did not. In
the latter cosmology, $a(t)\propto R_{\rm h}(t)$, and therefore
$R_{\rm ew}(t_{\rm ew})/R_{\rm h}(t_{\rm ew})=
R_{\rm ew}(t_0)/R_{\rm h}(t_0)$. In other words, the size of
the region with a uniform Higgs vev today is the same fraction
of $R_{\rm h}$ as it was at the EWPT, so the entire visible
Universe has a structure based on just a single Higgs vev 
manifested at the EWPT.

\section{Conclusion}
Wherever atomic and nuclear matter has been studied on cosmic
scales, no reliable evidence has ever been found of a breakdown
in their physical properties measured locally (see, e.g., ref.~\cite{Planck:2015}). 
For example, as far out as we can see in the Universe, all 
structures appear to be made out of matter, not antimatter. So if 
the baryon asymmetry is indeed due to the EWPT, its uniformity 
amplifies the argument of a uniform Higgs vev throughout the 
visible Universe. 

In this paper, we have shown that the EWPT may be avoided altogether
with an alternative choice of cosmology, specifically, the $R_{\rm h}=ct$
universe. We have demonstrated that, regardless of when an event
took place in the early Universe, the causally-connected region at
that time would have filled the entire visible Universe today. Therefore,
neither the phase transition associated with GUT 
(producing inflation), nor the electroweak phase transition due to 
the Higgs field being turned on, would have created observable
sub-horizon features in the $R_{\rm h}=ct$ cosmology.

We estimated the Hubble (gravitational) radius in this model assuming
the same time $t_{\rm ew}\sim 10^{-11}$ seconds for the EWPT as in the
standard model, and found that $R_{\rm h}(t_{\rm ew})\sim 0.3$ cm, much
smaller than the corresponding horizon size in $\Lambda$CDM. Even so,
the fact that $R_{\rm h}=ct$ had no early deceleration means that the 
causally-connected volume at that time would still have expanded sufficiently
to fill our entire visible Universe today. Of course, $t_{\rm ew}$ is likely
to be different in $R_{\rm h}=ct$ but, as we have noted, the actual time
at which the EWPT took place has no impact on the outcome. 

The EHP does not receive as much discussion today as does its GUT partner, 
perhaps because the latter is viewed as a more critical
ingredient of the standard model. But the reality is that the foundational
theory behind the Higgs mechanism for creating fermionic mass and separating
the electric and weak forces is now quite well established. If conflict
between the EWPT and cosmology remains unresolved, it is reasonable to
question the cosmological framework, as we have done in this paper. 
The $R_{\rm h}=ct$ model has been shown to fit many kinds of data better
than $\Lambda$CDM (see, e.g., Table~1 in ref.~\cite{Melia:2017b}). Indeed, there is
some evidence that the largely empirical, parametric formulation of
$\Lambda$CDM essentially produces optimized fits that mimic the predictions
in $R_{\rm h}=ct$ \cite{Melia:2015}. We believe that this alternative
model's ability to elegantly and simply avoid conceptual difficulties, such
as the CMB and EW horizon problems, is yet another strong argument in its favour.
 
{\acknowledgement
I acknowledge Amherst College for its support through a John Woodruff Simpson Fellowship. 
\endacknowledgement}

%=====================================================
%
%                           BIBLIOGRAPHY
%
%=====================================================

%=====================================================
\end{document}